\documentclass[conference]{IEEEtran}
\ifCLASSINFOpdf
     \usepackage[pdftex]{graphicx}
\usepackage{tikz}
\usetikzlibrary{shapes,snakes}
\usetikzlibrary{calc}
   \DeclareGraphicsExtensions{.pdf,.jpeg,.png}
\else
\fi
\usepackage{multicol}
\usepackage{multirow}
\usepackage{bigdelim}
\usepackage{amsfonts}
\usepackage{enumitem}
\usepackage{rotating}

\newcommand{\textcase}[2]
{
\setlength{\tabcolsep}{3pt}
\begin{tabular}{@{}l@{~~}llllll}
\ldelim\{ {#1}{1mm}
#2
\end{tabular}
}

\begin{document}
%
\title{On the Concept of Variable Roles and \\its Use in Software Analysis }



%
\author{\IEEEauthorblockN{Yulia Demyanova, 
Helmut Veith, 
Florian Zuleger
}
\IEEEauthorblockA{
Vienna University of Technology
}}


\maketitle
\begin{abstract}
Human written source code in imperative programming languages exhibits typical patterns for variable use such as flags, loop iterators, counters, indices, bitvectors etc. Although it is widely understood by practitioners that these variable roles are important for automated software analysis tools, they are not systematically studied by the formal methods community, and not well documented in the research literature. In this paper, we study the notion of variable roles on the example of basic types (int, float, char) in C. We propose a classification of the variables in a program by variable roles, and demonstrate that classical data flow analysis lends itself naturally both as a specification formalism and an analysis paradigm for this classification problem. We demonstrate the practical applicability of our method by predicting membership of source files to the different categories of the software verification competition SVCOMP 2013.
\end{abstract}
\IEEEpeerreviewmaketitle
\section{Introduction}
Programs written in imperative programming languages, (for example C, Java, Perl, Python etc.) share typical patterns of variable use, e.g. flags, loop iterators, counters, indices, bitvectors, temporary variables and so on. When recognising these patterns, a programmer has some expectations on how a variable can be used in the program, therefore we will call them \textit{variable roles}. For example, from a piece of C code \texttt{while(i$\textless$n) \{a[i++]=0;\}} one can deduce that \texttt{i} is a loop iterator and an array index, and from \texttt{x\&=y} that \texttt{x} is a bitvector.\\[3pt]
In standard systems like in the above mentioned languages,
there is no direct mapping from data types to roles - one type can correspond to one or more roles. For example, in C, the type \texttt{int} can be used to store such different values as boolean, file descriptor, bitvector and character literal. It is also not possible to explicitly define roles like loop iterator, array index, counter in these type systems. Moreover, one variable can have several roles, like \texttt{i} in the loop example above. In type systems, in contrast, one variable must be assigned one and only one type. Therefore, roles can not be regarded simply as refined types.
\\[3pt]
Information about variable roles is implicitly contained in the structure of the source code. One can often extract the role by syntactic analysis, for example by analysing statements of a given kind (e.g. matching array indices in array subscripts), or by looking for code patterns (e.g. \texttt{t=x; x=y; y=t;} is a typical pattern for temporary variable \texttt{t}).\\[3pt]
Importantly, variable roles are an informal and empirical notion -- i.e., they can be systematically studied and analysed, but they need to be treated as auxiliary heuristic information.
Thus, variable roles can {\em guide} a verification tool, but the {\em soundness} of a formal analysis must not depend on variable roles.
An important natural application is the use of variable roles
to create abstractions in software verification. For example, in C integer variables are used to store boolean flags, because there is no boolean type. However for such a variable \texttt{x} the predicate \texttt{x==0} is sufficient. Also when inferring invariants for an array, it is useful to know which variables are used as indexes for the array.
Unfortunately, most state-of-the-art verification papers treat a program as a logical formula and either ignore such implicit information, or treat it as undocumented heuristics. In this paper, we argue that these heuristics deserve a more systematic study.\\[3pt]
%
%
%
In this paper we define 14 variable roles with a standard data-flow analysis. Our definition serves at the same time as an algorithm to compute the roles. In order to choose the roles, we have manually investigated 5.2 KLOC of C code from cBench benchmark \cite{cBench} and assigned roles to basic type variables (int, float and char). When choosing the roles, we were inspired by typical programming patterns for variable use in real life programs. We have chosen the roles in such a way that a small number of roles is able to classify each occurring program variable in the programs we considered.\\[3pt]
As this short paper is reporting work in progress, we are currently exploring applications for variable roles. As mentioned earlier, variable roles can be used to create abstractions for software verification and to understand programs. We also conjecture that the frequency of variables of different roles in a program depends on the kind of the program, e.g. computationally intensive programs, device drivers or programs which extensively use dynamic data structures. We suggest that our method can be used to classify files (for example from benchmarks for different verification competitions) based on the number of variables of different roles. \\[3pt]
We have implemented a prototype tool, which maps basic-type (int, float, char) variables in C programs to sets of roles.
We then made the following experiment with the benchmarks from the software competition SVCOMP 2013 \cite{SVCOMP}. The competition distinguishes several categories of source files, e.g. device drivers, embedded systems, concurrent programs etc. This classification by human experts (who manually analysed and comprehended the source code) provides us with an excellent opportunity to benchmark our variable roles as follows:
With our tool we computed the frequency of different roles in each category and used this data to train a machine learning tool to predict the competition categories for new files. In a number of experiments we randomly selected a subset of the competition source files for training and used the remaining source files to check our prediction against human classification. The results of the experiments are encouraging - the prediction is successful in more than 80\% of the cases. These results are particularly impressive because our choice of the roles was based on examples from cBench rather than SVCOMP.
\begin{figure}[!t]
\centering
\begin{minipage}{\columnwidth}
\setlength{\columnsep}{-0.5 cm}
\begin{multicols}{2}
\begin{tabbing}
\hspace{0.5cm}\= \hspace{0.5cm}\= \hspace{0.5cm}\= \hspace{0.5cm}\=\kill
\scriptsize{1}\>int x, y, n = 0;\\
\scriptsize{2}\>\dots\\
\scriptsize{3}\>y=x;\\
\scriptsize{4}\>if (x) \{\\
\scriptsize{5}	\>\>do \{\\
\scriptsize{6}		\>\>\>n++; \\
\scriptsize{7}		\>\>\>x = x \&~(x-1);\\
\scriptsize{8}	\>\>\}\\
\scriptsize{9}\>\>while (x);\\
\scriptsize{10}\>\}\\
{\small(a) bitvector, counter, iterator}
\end{tabbing}
\begin{tabbing}
\hspace{0.5cm}\= \hspace{0.5cm}\= \hspace{0.2cm}\= \hspace{0.5cm}\=\kill
\scriptsize{1}\>int fd = open(path, flags);\\
\scriptsize{2}\>int c, val=0;\\
\scriptsize{3}\\
\scriptsize{4}\>while (read(fd, \&c, 1) \textgreater ~0 \\
\scriptsize{5}\>\>\&\& isdigit(c)) \\
\scriptsize{6}	\>\{\\
\scriptsize{7}	\>\>val = 10*val + c-'0');\\
\scriptsize{8}\>\}\\[24pt]
{\small(b) character, file descriptor, linear}
\end{tabbing}
\end{multicols}
\end{minipage}
\caption {Different patterns of use of integer variables}
\label{codesample}
\end{figure}
\subsection{Motivating examples}
Consider the C programs from Figure \ref{codesample}.
We will use these examples to informally introduce variable roles, formal definitions of which will be given in the next section.\\[3pt]
The program in Figure \ref{codesample}(a) calculates the number of non-zero bits in variable x. In every loop iteration a non-zero bit in x is set to zero and counter n is incremented. The loop continues until all bits are set to zero. Although variables x and n are declared of the same type int, however they are treated differently. For a human reading the program, statements \texttt{n=0} and \texttt{n=n+1} in the loop body signal that n is a counter (indeed, it is used to count the number of loop iterations). On the other hand, the value of variable x as an integer is not important for calculations, but rather individual bits in its binary representation matter. \\[3pt]
We define the roles by putting constraints on the operations in which a variable occurs. We require that a bitvector \textit{must} occur in at least one bitwise operation (bitwise AND, OR or XOR). For a counter variable  we require that it can only change its value in increment or decrement statements. 
Alternatively, it can be reset to zero. By analysing all assignments to the variable \texttt{n} (lines 1 and 6) we make sure that it satisfies these constrains.\\[3pt]
%
In real-life programs the use of variables can be ambiguous, when a variable which is used in one role for some time, is then used in an operation not typical for the role. For example consider a piece of C code \texttt{x\&=y; x<<1;}  where x is used as bitvector. Now for a slightly modified code \texttt{x\&=y; x*=2;} which is semantically  equivalent to the first one it is arguable whether x is a bitvector, because using a bitvector in an arithmetic operation in general does not make sense. We intentionally do not require that a bitvector does not occur in non-bitwise arithmetic operations (e.g. multiplication), and therefore with our definition x is assigned the role BITVECTOR in both cases. We consider such a definition of BITVECTOR reasonable, because during the manual investigation of a large code base we observed that in most cases variables used in bitwise operations are in fact treated as a collection of bits. An exception to this rule is bitwise shift, which is often used to replace multiplication by a power of two and therefore not used in our definition. Another argument is that since a variable can be assigned more than one role at the same time, we can consider the roles assigned to a variable altogether and choose the appropriate ones.\\[3pt]
The program in Figure \ref{codesample}(b) reads a decimal number from a text file and transforms it to numeric form. The result of computations is stored in variable val. In contrast, variables fd and c both take as their values output of functions, which are not part of the program: open() and read() respectively. The difference between the variables is that c is later used in calculations, while fd is only passed to function read() as a black box - its value does not directly affect the result of computations. One can make an assumption that c is a character, because it is passed as input to isdigit() function, which checks whether its parameter is a decimal digit character. Even though isdigit() is declared to take parameter of type int, the documentation says that c is "character to be checked, casted to int".\\[3pt]
We define character, file descriptor and linear roles as follows. We require that a character variable \textit{must} at least once be assigned either a character literal (e.g. \texttt{c='a'}) or another character variable, or used in one of the functions from the standard C library (e.g. \texttt{c=getchar()} or \texttt{isdigit(c)}). Again we do not restrict the operations in which a character \textit{can} occur, for example we allow that it is used in arithmetic operations, like a variable \texttt{c} in line 7. For a file descriptor we require that it is used at least once in a standard library function (e.g. \texttt{fd=open(path,flags)} or \texttt{read(fd,\&c,1)}). Finally, we require that a linear variable \textit{can} be assigned only linear combinations of linear variables, and both \texttt{c} and \texttt{val} satisfy this constraint.\\[6pt]
\textbf{Contributions}:
\begin{itemize}
\item We identify 14 variable roles that occur in practical programs.
\item We implement a prototype tool for C which assigns one or more roles to basic-type variables.
\item With our tool and machine learning techniques we predict the membership of a C programs in a competition category of the SW verificaiton competition SVCOMP 2013. We get encouraging results in a number of experiments.
\end{itemize}
\section{Formalisation of variable roles}
\subsection {Definition of the analysis}
We define variable roles with dataflow analysis as formulated in \cite{nielson1999principles}.
In the role definition we use our own language C$_{simpl}$, which represents a simplified version of C.
We do not specify its syntax and semantics due to the lack of space. For a program in C we translate every function to a program in C$_{simpl}$, and use this representation for intraprocedural analysis.\\[3pt]
%
%
%
In the definitions below we use the following notation: \textbf{Var} denotes the set of program variables, \textbf{Num} - all scalar constant literals (e.g. 0, 0.5, 'a') and \textbf{S}, \textbf{E} and \textbf{B} - the set of program statements, arithmetic and boolean expressions respectively. For the elements of these sets we use the same names in uncapitalised version (e.g. var for a program variable etc.).\\[3pt]
For a program s$\in \textbf{S}$ the result of analysis $R$ is computed using the function
$Res^R$, which is defined as follows: \\[6pt]
\[
	Res^R = Init^R \bigsqcup gen^R(s),
\]
%
where $Init^R \in \mathcal{P}(\textbf{Var})$ is the \textit{initial} set of variables, function $gen^R:\textbf{S} \cup \textbf{E} \cup \textbf{B} \rightarrow \mathcal{P}(\textbf{Var})$ maps every statement and expression to a set of \textit{generated} variables, and the sign $\bigsqcup$ is used as a placeholder for a set operation and must be instantiated for each analysis.\\[3pt]
Analysis $R$ is therefore defined with a tuple ($Init^R$, $\bigsqcup$, $gen^R$, c), where c$\in\{\textbf{f}, \textbf{o}\}$ indicates how to evaluate $Res^R$. When c is set to \textbf{f}, a fixed point of $Res^R$ is computed, i.e. $Res^R$ is iteratively recalculated until it does not change. When c is set to \textbf{o}, $Res^R$ is calculated in one iteration.\\
\begin{figure}
{\small
\begin{tabular}{@{}rcl}
\multicolumn{3}{c}{\textbf{BITVECTOR} \hspace{10pt}Init = $\emptyset$, $\bigsqcup$ = $\cup$, c = \textbf{O}}\\[6pt]
gen(var := e) &=&\hspace{-6pt}\textcase{2}{&\{var\} & if e ::= e$_1$ bitop e$_2$\\
	&$\emptyset$ & otherwise}\\
gen(if b then s$_1$ else s$_2$) &=& gen(b) $\cup$ gen(s$_1$) $\cup$ gen(s$_2$)\\
gen(s$_1$; s$_2$) &=& gen(s$_1$) $\cup$ gen(s$_2$)\\
gen(skip) &=& $\emptyset$\\
gen(while b do s) &=& gen(b) $\cup$ gen(s)\\[6pt]
gen(var) = gen(num) &=& $\emptyset$\\
gen(e$_1$ bitop e$_2$) &=& IsVar(e$_1$) $\cup$ IsVar(e$_2$) \\ &&$\cup$ gen(e$_1$) $\cup$ gen(e$_2$)\\
gen(e$_1$ aop e$_2$) &=& gen(e$_1$) $\cup$ gen(e$_2$)\\
gen(bitnot e) &=& IsVar(e) $\cup$ gen(e)\\[12pt]
\multicolumn{3}{c}{\textbf{LINEAR} \hspace{10pt}Init = \textbf{Var}, $\bigsqcup$ = $\setminus$, c = \textbf{F}}\\[3pt]
gen(x:=e) &=&\hspace{-6pt}\textcase{2}{&\{var\} & if lin(e)=false\\
& $\emptyset$ & otherwise}\\
gen(if b then s$_1$ else s$_2$) &=& gen(s$_1$) $\cup$ gen(s$_2$)\\
gen(s$_1$; s$_2$) &=& gen(s$_1$) $\cup$ gen(s$_2$)\\
gen(skip) &=& $\emptyset$\\
gen(while b do s) &=& gen(s)\\
gen(e) &=& $\emptyset$\\[6pt]
lin(num) &=& true\\
lin(var) &=&\hspace{-6pt}\textcase{2}{&true & if var $\in$ Res$^{LINEAR}$\\
&false & otherwise}\\
lin(e$_1$+e$_2$) &=& lin(e$_1$) $\land$ lin(e$_2$)\\
lin(e$_1$*e$_2$) &=&\hspace{-6pt}\textcase{3}{&lin(e$_2$) & if e$_1$ $\in$ \textbf{Num}\\
&lin(e$_1$) & if e$_2$ $\in$ \textbf{Num}\\
&false & otherwise}\\
lin(e$_1$ bitop e$_2$) &=& lin(bitnot e) = lin(e$_1$/e$_2$) = false
\end{tabular}
}
\caption{Formal definition of roles BITVECTOR and LINEAR}
\label{RoleDef}
\end{figure}
\subsection {Example of role definition}
In Figure \ref{RoleDef} we formally define the analysis for roles BITVECTOR and LINEAR\footnote{
The formal definition of all roles is given in the Appendix.}. An informal definition of the remaining roles is given in Table \ref{role_informal_def}. We will now show how the roles are computed on the example program from Figure \ref{codesample}(a), which is rewritten in $C_{simpl}$ in Figure \ref{coderewritten}(a).\\[3pt]
For the role BITVECTOR the analysis starts with an empty result ($Init=\emptyset$), the $\bigsqcup$ operation is defined to set union, and the result set is calculated in one iteration (c = \textbf{O}). When statement 2 is processed, variable x is added to the result set, because in this statement x is assigned the result of a bitwise AND operation. After that the result set does not change and evaluates (for the whole program) to \{x\}.\\[3pt]
The analysis for the role LINEAR is defined as a fixed point of the function $Res^R$ (c = \textbf{F}). It starts with the set \textbf{Var} of all program variables, i.e. \{x,y,n\}, and $\bigsqcup$ is defined to set minus. In the first iteration variable x is excluded from the result set at statement 2, because x is assigned a non-linear expression. In the second iteration variable y is excluded from the result set at statement 3, because it is assigned the value of x, and x does not belong to the result set. In the third iteration the result set does not change and the result of the anlaysis for the program evaluates to \{n\}.
\begin{table}[!t]
\caption{Informal definition of variable roles}
\label{role_informal_def}
\begin{center}
\vspace{-20pt}
\begin{tabular}{|@{~}p{2.2cm}|p{6cm}|}
\hline
\textbf{Role name} & \textbf{Informal definition}\\ 
\hline
SYNT\_CONST & is not assigned any value in the program\\
\hline
CONST\_ASSIGN & can be assigned only numeric literals or values of other variables which are assigned this role\\
\hline
COUNTER & can be assigned only in increment and decrement statements, or assigned zero\\
\hline
LINEAR & can be assigned only linear combinations of linear variables\\
\hline
BOOL & can be assigned zero, one, other boolean variables or boolean expression\\
\hline
INPUT & is passed to a function as a parameter by reference\\
\hline
BRANCH\_COND & must occur in the condition of if statement\\
\hline
BITVECTOR & must occur in a bitwise operation or assigned the result of a bitwise operation\\
\hline
UNRES\_ASSIGN & must be assigned the value of an array element\\
\hline
CHAR & can be assigned only character literals, values of other variables which are assigned this role, or initialised through a specific library function (e.g. \texttt{getchar()})\\
\hline
LOOP \_ITERATOR & must occur in the condition of the loop iterator and must be assigned in the loop body\\
\hline
FILE\_DESCR & must be assigned the result of a call to the library function \texttt{open()} or passed to \texttt{read()} or \texttt{write()} as a first parameter\\
\hline
ARRAY\_INDEX & must occur in an array subscript operation\\
\hline
\end{tabular}
\end{center}
\vspace{-20pt}
\end{table}
\begin{figure}[t!]
\centering
\begin{minipage}{\columnwidth}
\begin{multicols}{2}
\begin{tabbing}
\hspace{0.3cm}\= \hspace{0.4cm}\= \hspace{0.4cm}\= \hspace{0.5cm}\=\kill
n = 0; y =$^1$x;\\[3pt]
if (x != 0) \{ \\
	\> n = n + 1;\\
		\>x =$^{2}$ x bitand$^{3}$~(x - 1);\\[3pt]
		\>while$^{4}$ (x != 0) \{\\
		\>\>n = n + 1; \\
		\>\>x =$^{5}$ x bitand$^{6}$(x-1);\\
	\>\}\\
\}\\
{\small(a) code from fig. \ref{codesample}(a) in C$_{simpl}$}
\end{tabbing}
{\small
\begin{tabular}{|p{0.5cm}|p{1.2cm}|p{1.2cm}|}
\hline
\multicolumn{3}{|c|}{\textbf{BITVECTOR}}\\
\hline
\multicolumn{2}{|l|}{label} & gen(s)\\
\hline
\multicolumn{2}{|l|}{10,18} & \{x\}\\
\hline
\multicolumn{3}{|c|}{Init(vd)=$\emptyset$, Anal=\{x\}}\\
\hline
\multicolumn{3}{c}{ }\\
\hline
\multicolumn{3}{|c|}{\textbf{COUNTER}}\\
\hline
Iter. & label & gen(s)\\
\hline
1 & 9,17 & \{x\}\\
\hline
2 & 3 & \{y\}\\
\hline
\multicolumn{3}{|c|}{Init(vd)=\{x,y,n\}, Anal=\{n\}}\\
\hline
\multicolumn{3}{c}{}\\
\end{tabular}
}\\[6pt]
{\small(b) Result of the analysis}
\end{multicols}
\end{minipage}
\caption{Computation of roles}
\label{coderewritten}
\vspace{-5pt}
\end{figure}
\section{Implementation and experiments}
\begin{figure}[!t]
\setlength{\columnsep}{+3.5cm}
\begin{multicols}{2}
\begin{small}
\hbox{a) Relative numbers of roles for 2 categories}
\resizebox {2.2\columnwidth}{!}{
\begin{tikzpicture}
\coordinate (XAxisMin) at (0,0);
\coordinate (XAxisMax) at (10.5,0);
\coordinate (YAxisMin) at (0.5,-0.5);
\coordinate (YAxisMax) at (0.5,5.5);
\draw [thin, gray,-latex] (XAxisMin) -- (XAxisMax);
\draw [thin, gray,-latex] (YAxisMin) -- (YAxisMax);
\node[rotate=90, anchor=center] at (-0.8,3) {\Large{Relative numbers of roles, \%}};
\node[rotate=60, anchor=east] at (1,-0.5) {\large{SYNT\_CONST}};
\node[rotate=60, anchor=east] at (2,-0.5) {\large{CONST\_ASSIGN}};
\node[rotate=60, anchor=east] at (3,-0.5) {\large{COUNTER}};
\node[rotate=60, anchor=east] at (4,-0.5) {\large{LINEAR}};
\node[rotate=60, anchor=east] at (5,-0.5) {\large{BOOL}};
\node[rotate=60, anchor=east] at (6,-0.5) {\large{INPUT}};
\node[rotate=60, anchor=east] at (7,-0.5) {\large{BRANCH\_COND}};
\node[rotate=60, anchor=east] at (8,-0.5) {\large{BITVECTOR}};
\node[rotate=60, anchor=east] at (9,-0.5) {\large{UNRES\_ASSIGN}};
\node[rotate=60, anchor=east] at (10,-0.5) {\large{USED\_IN\_ARITHM}};
\node at (-0, 1) {\large{10\%}};
\node at (-0, 2) {\large{20\%}};
\node at (-0, 3) {\large{30\%}};
\node at (-0, 4) {\large{40\%}};
\node at (-0, 5) {\large{50\%}};
\node[draw,circle,inner sep=2pt,fill] at (1,6.97/10){};
\node[draw,circle,inner sep=2pt,fill] at (2,33.96/10){};
\node[draw,circle,inner sep=2pt,fill] at (3,46.08/10){};
\node[draw,circle,inner sep=2pt,fill] at (4,2.93/10){};
\node[draw,circle,inner sep=2pt,fill] at (5,51.93/10){};
\node[draw,circle,inner sep=2pt,fill] at (6,31.14/10){};
\node[draw,circle,inner sep=2pt,fill] at (7,40.36/10){};
\node[draw,circle,inner sep=2pt,fill] at (8,0/10){};
\node[draw,circle,inner sep=2pt,fill] at (9,0/10){};
\node[draw,circle,inner sep=2pt,fill] at (10,10.29/10){};
\node[draw,circle,inner sep=3pt] at (1,6.88/10){};
\node[draw,circle,inner sep=3pt] at (2,10.64/10){};
\node[draw,circle,inner sep=3pt] at (3,16.69/10){};
\node[draw,circle,inner sep=3pt] at (4,5.7/10){};
\node[draw,circle,inner sep=3pt] at (5,23.04/10){};
\node[draw,circle,inner sep=3pt] at (6,4.92/10){};
\node[draw,circle,inner sep=3pt] at (7,15.96/10){};
\node[draw,circle,inner sep=3pt] at (8,9.11/10){};
\node[draw,circle,inner sep=3pt] at (9,9.06/10){};
\node[draw,circle,inner sep=3pt] at (10,0/10){};
\node[draw,circle,inner sep=2pt,fill] at (1,67/10){};
\node [anchor=west] at (1.5, 67/10){\Large{Control Flow and Integer Variables}};
\node[draw,circle,inner sep=2pt] at (1,62/10){};
\node [anchor=west] at (1.5, 62/10){\Large{Linux Device Drivers 64-bit}};
\foreach \x in {1, 2,...,10}{
\draw[thin, gray, -latex] (\x, -0.1) -- (\x, 0.1);
}
\foreach \y in {1, 2,...,5}{
\draw[thin, gray, -latex] (0.5+-0.1, \y) -- (0.5+0.1, \y);
}

  \end{tikzpicture}
}
\newpage
b) Prediction error \\
in different settings
\\[3pt]
\begin{tabular}{@{}c|@{}c|@{}c}
\hline
\begin{sideways}{\parbox{1.6cm}{Training set, \% of all files}}\end{sideways} & \begin{sideways}{\parbox{1.6cm}{Error of choice 1}}\end{sideways} & \begin{sideways}{\parbox{1.6cm}{Error of choice 2}}\end{sideways}\\[3pt]
\hline
90\% & 15.94\% & 2.90\%\\[3pt]
80\% & 14.81\% & 5.93\%\\[3pt]
70\% & 16.20\% & 7.98\%\\[3pt]
60\% & 19.77\% & 7.98\%\\[3pt]
50\% & 18.60\% & 8.54\%\\
\end{tabular}
\end{small}
\end{multicols}
\vspace{-20pt}
\caption{Comparison of categories and automatic classification of files}
\vspace{-20pt}
\label{cross_check_exp}
\end{figure}
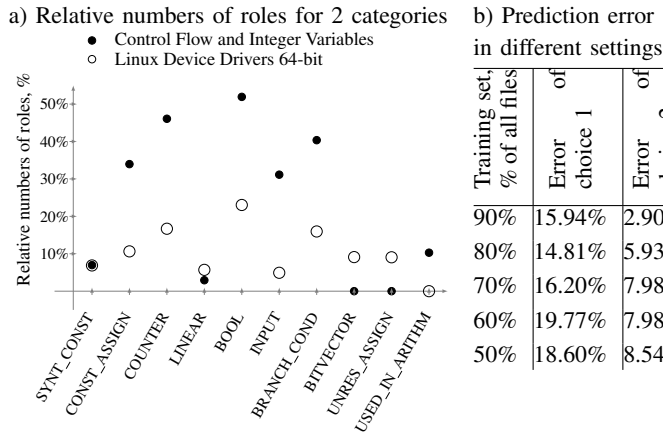

We implemented a prototype tool using clang compiler. We  handle pointers and function calls as follows: when a variable is assigned a pointer dereference (e.g. \texttt{n=*ptr}, \texttt{n=arr[i]}), or passed to a function as a parameter by reference, the result set does not change. To find a trade-off between safe and accurate analysis, we signal about such situations by assigning the variable the role "unresolved assignment". For some roles, namely char and file descriptor, we
use the information about function calls,
and search for the calls of the functions from the standard library (e.g. \texttt{open}, \texttt{getchar} etc).\\[3pt]
%
We compare the relative numbers of roles as shown in Figure \ref{cross_check_exp}a) for categories "Control Flow and Integer Variables" and "Linux Device Drivers". As expected, we observe that boolean flags and branching operations as well as counters, arithmetic operations and constant assignments are typical for the first category, while Linux drivers extensively use bitvectors and pointers.\\[3pt]
In our experiments we used a multiclass vector support machine \cite{weston1998multi} to predict the categories for files with some probability, e.g. the probability that the file is a driver is 60\%, and that it is a concurrent program - 35\% and so on. We translated the relative numbers of roles into the input format of a machine learning tool (http://www.cs.waikato.ac.nz/ml/weka) as follows: each source file represented one training example with the category corresponding to the class, and relative numbers of roles representing the vector of float attributes.\\[3pt]
We ran the experiments for the sizes of the training sets from 90\% to 50\% of all files and obtained the prediction error of approximately 15\% as shown in Figure \ref{cross_check_exp}b). When we analysed the second most probable choices, we observed the error of appr. 7\%.\\
%
\section{Related work}
The term variable roles was inspired by \cite{Sajaniemi:2002:EAR:795687.797809}, which informally defines roles as \textit{patterns of how variables are initialised and updated}. The authors have defined nine roles, implemented a tool for assigning roles to variables using static analysis and evaluated it on Pascal programs from textbooks. The work leaves open the question of formalising the notion of variable roles as well as of the possibility of applying the method to real-word programs.\\[3pt]
%
\cite{Engler:2001:BDB:502059.502041} uses implicit knowledge
in the form of programmer's \textit{beliefs}, i.e. propositional statements about program variables and functions, for bug finding. The authors use static analysis to extract must (e.g. "a pointer is not null")
and may (e.g. "calls to functions \begin{it}{f()}\end{it} and \begin{it}{g()}\end{it} should be paired") statements. 
Since the project has grown into a commercial tool (Coverity), publicly available research results have been limited.\\[3pt]
%
In \cite{Rondon:2008:LT:1375581.1375602} predicate abstraction over a fixed set of predicates is used to infer so called \textit{liquid} types, i.e. refinement of types with a conjunction of propositional predicates (e.g. $x \textgreater 0 \land x \textless 5$). We consider this approach to be complementary to ours, because it does not use any information from the source code other than the transition relation, and concentrates on arithmetic properties of variables.\\[3pt]
%
Variable names and comments as an additional source of knowledge about a program
have been systematically studied in program comprehension.
The \textit{Latent Semantic Indexing} technique \cite{Deerwester90indexingby} allows to query on the program source code using words in natural language, and to obtain a list of functions ranked with a \textit{similarity characteristic}. The latter is calculated from the number of occurrences of the words from the query in variable names and comments of a function. The rules for naming variables in real-word programs are studied in \cite{deissenboeck2006concise}, and \cite{lawrie2007extracting} suggests a method for expanding abbreviated identifiers to full words. We regard using these techniques in our approach as future work. 

\section*{Acknowledgment}
This work received funding in part by the Austrian National Research Network S11403-N23 (RiSE) of the Austrian Science Fund (FWF) and by the Vienna Science and Technology Fund (WWTF) grant PROSEED.
%
\bibliographystyle{IEEEtran}
\bibliography{source.bbl}
\onecolumn
\begin{figure}
{\small
\begin{tabular}{rcl}
\multicolumn{3}{c}{\normalsize{APPENDIX}}\\[12pt]
\multicolumn{3}{c}{\normalsize{\textbf{Syntax of the language C$_{simpl}$}}}\\[12pt]
\textbf{E} &::=& var $\mid$ num $\mid$ E~ aop~ E $\mid$ E~ bitop~ E  $\mid$ bitnot E $\mid$ avar[E]\\
aop &::=& + $\mid$ - $\mid$ * $\mid$ / \\
bitop &::=& bitor $\mid$ bitand $\mid$ bitxor\\
\textbf{B} &::=& E~ compop~ E $\mid$ not~ B $\mid$ B ~logop~ B\\
compop &::=& = $\mid$ != $\mid$ $\textgreater$ $\mid$ $\textless$ $\mid$ $\geq$ $\mid$ $\leq$ \\
logop &::=& $\land$ $\mid$ $\lor$\\
\textbf{S} &::=& var := E $\mid$ avar[E] := E $\mid$ if B then S else S $\mid$ S;S $\mid$ skip $\mid$ while B do S $\mid$ call p(Par)\\
Par &::=& E $\mid$ Par, E\\
P &::=& proc p (VD) begin VD C end $\mid$ Proc;Proc $\mid$ $\varepsilon_P$ \\
VD &::=& var x $\mid$ VD; VD $\mid$ $\varepsilon_{VD}$\\
Prog &::=& begin VD Proc end\\[6pt]
\multicolumn{3}{l}{ Note: 1) the first parameter of a function is the returned value}\\
\multicolumn{3}{l}{ ~~~~~~~2) avar is an array variable}\\[6pt]
\end{tabular}
}
\caption{Syntax of C$_{simpl}$}
\label{SyntaxCsimpl}
\end{figure}
\begin{figure}[!t]
{\small
\begin{tabular}{rcl}
\multicolumn{3}{c}{\normalsize{\textbf{Definition of variable roles}}}\\[12pt]
IsVar(var) &=& \{var\}\\
IsVar(num) = &=& $\emptyset$\\
IsVar(e1 aop e2) &=& IsVar(e1 bitop e2) = IsVar(bitnot e) = IsVar(b1 logop b2) = IsVar(not b) = IsVar(avar[e]) = $\emptyset$\\
IsVar(s) &=& $\emptyset$\\[6pt]
\multicolumn{3}{c}{\textbf{One-run positive roles}: $Init^R=\emptyset$, $\bigsqcup=\cup$, c=\textbf{O}}\\[6pt]
\textbf{BITVECTOR}\\
gen(var := e) &=& gen(e) $\cup$\textcase{2}{&\{var\} & if e ::= e1 bitop e2\\
	&$\emptyset$ & otherwise}\\
gen(avar[e1] := e2) &=& gen(ax[e1]) $\cup$ gen(e2)\\
gen(if b then s1 else s2) &=& gen(b) $\cup$ gen(s1) $\cup$ gen(s2)\\
gen(s1; s2) &=& gen(s1) $\cup$ gen(s2)\\
gen(skip) &=& $\emptyset$\\
gen(while b do s) &=& gen(b) $\cup$ gen(s)\\
gen(call p(e1, ..., en)) &=& $\bigcup\limits_{1 \leq i \leq n}$ gen(ei)\\[6pt]
gen(var) = gen(num) &=& $\emptyset$\\
gen(e1 bitop e2) &=& IsVar(e1) $\cup$ IsVar(e2) $\cup$ gen(e1) $\cup$ gen(e2)\\
gen(e1 aop e2) &=& gen(e1) $\cup$ gen(e2)\\
gen(bitnot e) &=& IsVar(e) $\cup$ gen(e)\\
gen(b1 logop b2) &=& gen(b1) $\cup$ gen(b2)\\
gen(not b) &=& gen(b)\\
gen(avar[e]) &=&gen(e)\\[12pt]
\textbf{FILE\_DESCR}\\
gen(var := e) &=& gen(ax[e1] := e2) = $\emptyset$\\
gen(if b then s1 else s2) &=& gen(s1) $\cup$ gen(s2)\\
gen(s1; s2) &=& gen(s1) $\cup$ gen(s2)\\
gen(skip) &=& $\emptyset$\\
gen(while b do s) &=& gen(s)\\
gen(call p(e1, ..., en)) &=& \textcase{3}{&IsVar(e1) & if p ::= open and and 3$\leq$n$\leq$4\\
&IsVar(e2) & if (p ::= read or p ::= write) and en=4\\
&$\emptyset$ & otherwise}\\
gen(e) = gen(b) &=& $\emptyset$\\[8pt]
%
\end{tabular}
}
\caption{One run-positive roles}
\label{FullRoleDef}
\end{figure}
\clearpage

\begin{figure}[t!]
{\small
\begin{tabular}{rcl}
\multicolumn{3}{c}{\textbf{One-run positive roles} (cont.)}\\[6pt]
\textbf{ARRAY\_INDEX}\\
gen(var := e) &=& gen(e)\\
gen(avar[e1] := e2) &=& IsVar(e1) $\cup$ gen(e2)\\
gen(if b then s1 else s2) &=& gen(b) $\cup$ gen(s1) $\cup$ gen(s2)\\
gen(s1; s2) &=& gen(s1) $\cup$ gen(s2)\\
gen(skip) &=& $\emptyset$\\
gen(while b do s) &=& gen(b) $\cup$ gen(s)\\
gen(call p(e1, ..., en)) &=& $\bigcup\limits_{1 \leq i \leq n}$gen(ei)\\
gen(avar[e]) &=& IsVar(e)\\
gen(e1 aop e2) &=& gen(e1) $\cup$ gen(e2)\\
gen(e1 bitop e2) &=& gen(e1) $\cup$ gen(e2)\\
gen(bitnot e) &=& gen(e)\\
gen(b1 logop b2) &=& gen(b1) $\cup$ gen(b2) \\
gen(not b) &=& gen(b) \\[8pt]
\textbf{ARRAY\_SIZE}\\
gen(var := e) &=& $\emptyset$\\
gen(avar[e1] := e2) &=& $\emptyset$\\
gen(if b then s1 else s2) &=& gen(s1) $\cup$ gen(s2)\\
gen(s1; s2) &=& gen(s1) $\cup$ gen(s2)\\
gen(skip) &=& $\emptyset$\\
gen(while b do s) &=& gen(s)\\
gen(call p(e1, ..., en)) &=&\textcase{2}{&IsVar(e2) & if p ::= malloc and and n=2\\
&$\emptyset$ & otherwise}\\
gen(e) = gen(b) &=& $\emptyset$\\[8pt]
\textbf{UNRES\_ASSIGN}\\
gen(var := e) &=&\textcase{2}{&\{var\} & if e ::= ax[e1]\\
& $\emptyset$ & otherwise}\\
gen(avar[e1] := e2) &=& $\emptyset$\\
gen(if b then s1 else s2) &=& gen(s1) $\cup$ gen(s2)\\
gen(s1; s2) &=& gen(s1) $\cup$ gen(s2)\\
gen(skip) &=& $\emptyset$\\
gen(while b do s) &=& gen(s)\\
gen(call p(e1, ..., en)) &=&$\emptyset$\\
gen(e) = gen(b) &=& $\emptyset$\\[12pt]
textbf{OUTPUT}\\
gen(var := e) = gen(avar[e1] := e2) &=& $\emptyset$\\
gen(if b then s1 else s2) &=& gen(s1) $\cup$ gen(s2)\\
gen(s1; s2) &=& gen(s1) $\cup$ gen(s2)\\
gen(skip) &=& $\emptyset$\\
gen(while b do s) &=& gen(s)\\
gen(call p(e1, ..., en)) &=&\textcase{5}{&$\bigcup\limits_{2 \leq i \leq n}$ IsVar(ei) & if p ::= printf\\
&$\bigcup\limits_{3 \leq i \leq n}$ IsVar(ei) & if (p ::= sprintf or p ::= fprintf)\\
&$\emptyset$ & otherwise}\\
gen(e) = gen(b) &=& $\emptyset$\\[12pt]
\textbf{INPUT}\\
gen(var := e) = gen(avar[e1] := e2) &=& $\emptyset$\\
gen(if b then s1 else s2) &=& gen(s1) $\cup$ gen(s2)\\
gen(s1; s2) &=& gen(s1) $\cup$ gen(s2)\\
gen(skip) &=& $\emptyset$\\
gen(while b do s) &=& gen(s)\\
gen(call p(e1, ..., en)) &=&$\bigcup\limits_i$ IsVar(ei) where i-th parameter is passed by reference to p\\
gen(e) = gen(b)&=& $\emptyset$\\[12pt]
\end{tabular}
}
\caption{One run-positive roles (cont.)}
\label{OneRunPosCont}
\end{figure}

\begin{figure}[t!]
{\small
\begin{tabular}{rcl}
\multicolumn{3}{c}{\textbf{One-run positive roles} (cont.)}\\[6pt]
\textbf{BRANCH\_COND}\\
gen(var := e) = gen(avar[e1] := e2) &=& $\emptyset$\\
gen(if b then s1 else s2) &=& vars(b)\\
gen(s1; s2) &=& gen(s1) $\cup$ gen(s2)\\
gen(skip) &=& $\emptyset$\\
gen(while b do s) &=& gen(s)\\
gen(call p(e1, ..., en)) &=&$\emptyset$\\
gen(e) = gen(b) &=& $\emptyset$\\
vars(num) &=& $\emptyset$\\
vars(var) &=& \{var\}\\
vars(b1 logop b2) &=& vars(b1) $\cup$ vars(b2)\\
vars(not b) &=& vars(b)\\
vars(e1 compop e2)  = vars(e1 aop e2)  &=& vars(e1 bitop e2) = vars(e1) $\cup$ vars(e2)\\
vars(bitnot e) &=& vars(e)\\[12pt]
\textbf{USED\_IN\_ARITHM}\\
gen(var := e) &=& gen(e)\\
gen(avar[e1] := e2) &=& gen(e1) $\cup$ gen(e2)\\
gen(if b then s1 else s2) &=& gen(b) $\cup$ gen(s1) $\cup$  gen(s2)\\
gen(s1; s2) &=& gen(s1) $\cup$ gen(s2)\\
gen(skip) &=& $\emptyset$\\
gen(while b do s) &=& gen(b) $\cup$  gen(s)\\
gen(call p(e1, ..., en)) &=&$\bigcup\limits_{1 \leq i \leq n}$gen(ei) \\
gen(e1 aop e2) &=& IsVar(e1) $\cup$ IsVar(e2) $\cup$ gen(e1) $\cup$ gen(e2)\\
gen(e1 bitop e2) &=& gen(e1) $\cup$ gen(e2)\\
gen(bitnot e) &=& gen(e)\\
gen(b1 logop b2) &=& gen(b1) $\cup$ gen(b2)\\
gen(not b) &=& gen(b)\\[12pt]
\textbf{LOOP\_IT}\\
gen(var := e) = gen(avar[e1] := e2) &=& $\emptyset$\\
gen(if b then s1 else s2) &=& gen(s1) $\cup$ gen(s2)\\
gen(s1; s2) &=& gen(s1) $\cup$ gen(s2)\\
gen(skip) &=& $\emptyset$\\
gen(while b do s) &=& varsB(b) $\cap$ varsS(s)\\
gen(call p(e1, ..., en)) &=& $\emptyset$\\
gen(e) =gen(b) &=& $\emptyset$\\[6pt]
varsB(e1 compop e2) &=& IsVar(e1) $\cup$ IsVar(e2)\\
varsB(b1 logop b2) &=& varsB(b1) $\cup$ varsB(b2)\\
varsB(not b) &=& varsB(b)\\[6pt]
varsS(var:=e) &=& \{var\}\\
varsS(if b then s1 else s2) &=& varsS(s1) $\cup$ varsS(s2)\\
varsS(s1; s2) &=& varsS(s1) $\cup$ varsS(s2)\\
varsS(skip) &=& $\emptyset$\\
varsS(while b do s) &=& varsS(s)\\
varsS(call p(e1, ..., en)) &=& $\emptyset$\\[6pt]
\end{tabular}
}
\caption{One run-positive roles (cont.)}
\label{OneRunPosCont}
\end{figure}

\begin{figure}
{\small
\begin{tabular}{rcl}
\multicolumn{3}{c}{\textbf{Fixed-point negative roles}: $Init^R=\textbf{Vars}, \bigsqcup=\setminus$, c = \textbf{F}}\\[6pt]
\textbf{LINEAR}\\
gen(var:=e) &=&\textcase{2}{&\{var\} & if lin(e)=false\\
& $\emptyset$ & otherwise}\\
gen(avar[e1] := e2) &=& $\emptyset$\\ 
gen(if b then s1 else s2) &=& gen(s1) $\cup$ gen(s2)\\
gen(s1; s2) &=& gen(s1) $\cup$ gen(s2)\\
gen(skip) &=& $\emptyset$\\
gen(while b do s) &=& gen(s)\\
gen(call p(e1, ..., en)) &=& $\emptyset$\\
gen(e) = gen(b) &=& $\emptyset$\\[6pt]
lin(num) &=& true\\
lin(var) &=&\textcase{2}{&true & if var $\in$ $Res^{LINEAR}$\\
&false & otherwise}\\
lin(e1+e2) &=& lin(e1) $\land$ lin(e2)\\
lin(e1*e2) &=&\textcase{3}{&lin(e2) & if e1$\in\textbf{Num}$\\
&lin(e1) & if e2$\in\textbf{Num}$\\
&true & otherwise}\\
lin(e1 bitop e2) &=& lin(bitnot e) = lin(e1/e2) = false\\[12pt]
\textbf{COUNTER}\\
gen(var:=e) &=& \textcase{2}{& $\emptyset$ & if e$\in\textbf{Num}$ and e::=0\\
 &\{var\} $\setminus$ sumd(e) & otherwise}\\
gen(avar[e1] := e2) &=& $\emptyset$\\ 
gen(if b then s1 else s2) &=& gen(s1) $\cup$ gen(s2)\\
gen(s1; s2) &=& gen(s1) $\cup$ gen(s2)\\
gen(skip) &=& $\emptyset$\\
gen(while b do s) &=& gen(s)\\
gen(call p(..., modi ei, ...)) &=& $\emptyset$\\
gen(e) = gen(b) &=& $\emptyset$\\[6pt]
sumd(num) &=& $\emptyset$\\
sumd(var) &=& \{var\}\\
sumd(e1+e2) &=& \textcase{3}{&IsVar(e1) & if e2$\in\textbf{Num}$\\
& IsVar(e2) & if e1$\in \textbf{Num}$\\
&$\emptyset$ & otherwise\\}\\
sumd(e1-e2) &=& \textcase{2}{&IsVar(e1) & if e2$\in\textbf{Num}$\\
&$\emptyset$ & otherwise\\}\\
sumd(e1*e2) &=& sumd(e1/e2) = sumd(e1 bitop e2) = sumd(bitnot e) = $\emptyset$\\[12pt]
\textbf{CONST\_ASSIGN}\\
gen(var:=e) &=& \textcase{2}{& $\emptyset$ & if isConst(e)=true\\
 &\{var\} & otherwise}\\
gen(avar[e1] := e2) &=& $\emptyset$\\ 
gen(if b then s1 else s2) &=& gen(s1) $\cup$ gen(s2)\\
gen(s1; s2) &=& gen(s1) $\cup$ gen(s2)\\
gen(skip) &=& $\emptyset$\\
gen(while b do s) &=& gen(s)\\
gen(call p(..., modi ei, ...)) &=& $\emptyset$\\
gen(e) = gen(b) &=& $\emptyset$\\[6pt]
isConst(num) &=& $\emptyset$\\
isConst(var) &=&\textcase{2}{&true & if var$\in Res^{CONST\_ASSIGN}$\\
&$\emptyset$ & otherwise}\\
isConst(e1 aop e2) &=& isConst(e1) $\land$ isConst(e2)\\
isConst(e1 bitop e2) &=& isConst(e1) $\land$ isConst(e2)\\
isConst(bitnot e) &=& isConst(e)\\[12pt]
\end{tabular}
}
\caption{Fixed-point negative roles}
\label{FullRoleDef}
\end{figure}
\begin{figure}
{\small
\begin{tabular}{rcl}
\multicolumn{3}{c}{\textbf{Fixed-point negative roles} (cont.)}\\[6pt]
\textbf{BOOL}\\
gen(var:=e) &=&\textcase{2}{& $\emptyset$ & if isBool(e)=true \\
&\{var\} & otherwise}\\
gen(avar[e1]:=e2) &=&gen(e1) $\cup$ gen(e2)\\
gen(if b then s1 else s2) &=& gen(b) $\cup$ gen(s1) $\cup$ gen(s2)\\
gen(s1; s2) &=& gen(s1) $\cup$ gen(s2)\\
gen(skip) &=& $\emptyset$\\
gen(while b do s) &=& gen(b) $\cup$ gen(s)\\
gen(call p(e1, ..., en)) &=& $\bigcup\limits_{1 \leq i \leq n}$gen(ei)\\
gen(var) = gen(num) &=& $\emptyset$\\
gen(e1 aop e2) = &=& IsVar(e1) $\cup$ IsVar(e2)\\
gen(e1 bitop e2) &=& IsVar(e1) $\cup$ IsVar(e2)\\
gen(bitnot e) &=& IsVar(e)\\[6pt]
isBool(var) &=&\textcase{2}{&true & if var $\in Res^{BOOLEAN}$\\
&false & otherwise} \\
isBool(num) &=&\textcase{2}{&true & if num::=0 or num::=1\\
&false & otherwise}\\
isBool(b) &=& true\\
isBool(e1 aop e2) &=& false\\
isBool(e1 bitop e2) &=& false\\
isBool(bitnot e) &=& false\\
\end{tabular}
}
\caption{Fixed-point negative roles (cont.)}
\label{FullRoleDef}
\end{figure}
\begin{figure}
{\small
\begin{tabular}{rcl}
\multicolumn{3}{c}{\textbf{Fixed-point positive roles}: $Init^R=\emptyset, \bigsqcup=\cup$, c = \textbf{F}}\\[6pt]
\textbf{CHAR}\\
gen(var:=e) &=& \textcase{2}{&\{var\} & if isChar(e)=true\\
&$\emptyset$ & otherwise} \\
gen(avar[e1]:=e2) &=& $\emptyset$\\
gen(if b then s1 else s2) &=& gen(s1) $\cup$ gen(s2)\\
gen(s1; s2) &=& gen(s1) $\cup$ gen(s2)\\
gen(skip) &=& $\emptyset$\\
gen(while b do s) &=& gen(s)\\
gen(call p(e1, ..., en)) &=& \textcase{8}{&IsVar(e1) & if (n=1 and p ::= getchar)\\
& & or (n=2 and (p ::= getc or p ::= fgetc or p ::= tolower or p ::= toupper))\\[6pt]
& IsVar(e2) & if (n=2 and (p::=putchar or p::=tolower\\
&&or p::=toupper or p::=isalnum or p::=isblank or p::=iscntrl\\
&&or p::=isdigit or p::=isgraph or p::=islower or p::=isprint\\
&&or p::=isputnct or p::=isspace or p::=isupper or p::=isxdigit))\\
& $\emptyset$ & otherwise}\\[6pt]
gen(e) = gen(b) &=& $\emptyset$\\[6pt]
isChar(num) &=& \textcase{2}{&true & if num is character literal\\
&false & otherwise}\\
isChar(var) &=& \textcase{2}{&true & if var $\in Res^{CHAR}$\\
&false & otherwise}\\
isChar(e1 aop e2) &=& isChar(e1 bitop e2) = isChar(bitnot e) = isChar(b) =  $\emptyset$\\
\end{tabular}
}
\caption{Fixed-point positive roles}
\label{FullRoleDef}
\end{figure}
\begin{figure}
{\small
\begin{tabular}{rcl}
\multicolumn{3}{c}{\textbf{One-run negative roles}: $Init^R=\textbf{Vars}, \bigsqcup=\setminus$, c = \textbf{O}}\\[6pt]
\textbf{SYNT\_CONST}\\
gen(var:=e) &=&\{var\}\\
gen(avar[e1]:=e2) &=&$\emptyset$\\
gen(if b then s1 else s2) &=& gen(s1) $\cup$ gen(s2)\\
gen(s1; s2) &=& gen(s1) $\cup$ gen(s2)\\
gen(skip) &=& $\emptyset$\\
gen(while b do s) &=& gen(s)\\
gen(call p(e1, ..., en)) &=& $\emptyset$\\
gen(e) = gen(b) &=& $\emptyset$\\[6pt]
\end{tabular}
}
\caption{One-run negative roles}
\label{FullRoleDef}
\end{figure}
\end{document}